# Uplink Scheduling for LTE Video Surveillance Systems

Yen-Kai Liao[1], Chih-Hang Wang[1], De-Nian Yang[2], Wen-Tsuen Chen[2]
[1]Department of Computer Science, National Tsing Hua University, Hsin-Chu, 300, Taiwan
[2]Institute of Information Science, Academia Sinica, Nankang, Taipei 115, Taiwan
E-mail:jack32688@gmail.com, superwch7805@gmail.com, dnyang@iis.sinica.edu.tw, and chenwt@iis.sinica.edu.tw

*Abstract*—Due to the proliferation of applications for the Internet of Things, an increasing number of machine to machine (M2M) devices are being deployed. In particular, one of the M2M applications, video surveillance, has been widely discussed. Long Term Evolution (LTE), which can provide a high rate of data transmission and wide range of coverage, is a promising standard to serve as an M2M video surveillance system. In this paper, we studied a performance maximization problem in an LTE video surveillance system. Given a set of objects and a set of cameras, each camera has its own performance grade and its own coverage. The goal is to maximize the performance of the system by allocating limited resources to cameras while all objects should be monitored by the selected cameras. We propose a heuristic method to select the cameras and allocate resources to them to solve the problem. Moreover, to reduce the load of the LTE system, a dynamic adjustment method is also proposed.

## I. INTRODUCTION

To meet the increasing demand for mobile wireless access, the Third Generation Partnership Project (3GPP) Long Term Evolution (LTE) has attracted significant attention. LTE provides high network capacity, high transmission speed, and a wide coverage range. Orthogonal Frequency-Division Multiple Access (OFDMA) is adopted in LTE downlink (DL) transmission while Single-carrier FDMA (SC-FDMA) is adopted in the LTE uplink (UL) transmission to conserve power. The scheduling algorithm of resource allocation plays an important role for LTE, since an efficient algorithm can greatly increase the LTE system throughput. In the LTE DL system, due to the multi-user diversity gain, most of the algorithms use the channel conditions as the comparison metric to maximize the aggregate throughput. Some previous works modified the well-known Proportional Fair scheduler to ensure the fairness among the users [1]. Some other works took the quality of service (QoS) of each user into consideration [2]. In such algorithms, users with higher priority will be served earlier. Differing from the LTE DL system, the allocation policy used by the scheduler in the LTE UL must consider the contiguity constraint of SC-FDMA. Contiguity constraint means one user must be allocated adjacent resource blocks (RBs). Thus, although the comparison criteria of scheduling for the LTE UL are similar to those of LTE DL, the approaches are different. Most schedulers for UL have considered information related to adjacent RBs while allocating resources to users. However, neither for DL nor for UL, there is no scheduler which takes into consideration the demand for a specific service.

Machine to machine (M2M) communications, also called machine type communications (MTC), is a special type of communication between devices that communicates without human intervention. In recent years, M2M communication has been standardized in LTE. Due to its characteristics of low cost, low energy consumption, and wide range of the applications, it can be seen that M2M devices will be widely deployed in the near future such as surveillance cameras for city monitoring.

The video surveillance system is one promising M2M application. Recently, due to major terrorist attacks and criminal events, the demand for sophisticated surveillance systems is increasing. In the past, to carry the high load required for the surveillance systems, wired network or IEEE 802.11 network was used to transmit multimedia data. Nowadays, in contrast, the video surveillance system can be applied in an LTE network due to its high capacity. For example, police departments have collaborated with wireless ISPs to roll out innovative surveillance systems on 3G/4G networks [21], [22].To support the LTE video surveillance application, in the markets, there are already surveillance devices that can communicate through LTE [24]. Without using cables, the deployment of the application can have a lower cost and the system can be maintained more easily.

However, the bandwidth consumption for a video surveillance system is still a challenge for LTE when serving large number of devices. In such monitoring applications, the different data may have correlations in two domains: spatial and temporal. Spatial correlation means that an object may be covered by different cameras and the temporal correlation means that a view may not substantially change over a short period. Our goal is to use the correlations between different data to reduce the load of applying video surveillance in an LTE network while reaching a certain level of surveillance. Within the limited bandwidth in LTE, efficient scheduling is needed to avoid congestion and improve the system capacity. In the literature, camera selection and placement problems in surveillance systems have been widely studied. Using a minimum number of cameras to cover all the objects of interest is the main issue of such problems. However, when taking channel condition into consideration, such results may not be the best choice. A coverage problem with minimum wireless resources is another issue. Although the results of both problems can cover all the objects of interest, some objects may not be well captured in practice. Moreover, the network environment may be able to support more cameras to improve the performance of the system.

In this work, we assume that each camera can only monitor objects within a fixed coverage range and that each camera will upload the video in real time. According to the amount and the view angles of objects monitored, the view taken by each camera has its own grade. With these grades, we can know the performance of each camera. We take channel condition and the performance of each camera into consideration to decide whether or not to allocate resources to a camera. Given a surveillance system which includes a set of objects and a set of cameras, our goal is to maximize the monitoring performance of the system under the limited resources while having all objects ideally being covered by at least one camera.

The remainder of this paper is organized as follows. Section II discusses related work in three aspects. Section III introduces the backgrounds and the problem formulations. Section IV describes our methods. Section V presents the simulation results and their discussion. Finally, conclusions are offered about this work in Section VI.

## II. Related Work

### A. Data gathering in M2M application

Over the past few years, the concepts of a smart city and smart home have attracted the attention of many people. To make such applications a reality, M2M devices are being deployed extensively. Since the amount of M2M devices is large and the spectrum resources are limited, how to gather data from M2M devices efficiently has become a significant and pressing issue. Many previous works have studied this problem [3], [4]. Fu et al. [3] designed a centralized reporting mechanism and a distributed reporting mechanism for multi-type real-time monitoring based on the concept of sensing region and data validity in M2M networks. In [4], the authors defined "useful" values for devices based on entropy and used these values to optimize the system performance instead of maximizing the number of machines. Most of the approaches for the M2M data gathering problem aim to minimize the amount of sensor nodes in order to ease the network congestion and save power for the M2M devices. However, rather than minimizing the amount of sensor nodes, we try to maximize the monitoring performance under the coverage constraint for an LTE surveillance system.

### B. LTE resource allocation

LTE uplink resource scheduling has been discussed widely [5], [6], [7], [8], [9]. Although the smallest unit for uplink scheduling, the RB, is the same as for LTE downlink scheduling, there is one significantly different constraint in the LTE uplink, where the contiguity constraint must be fulfilled. Thus, a scheduling algorithm for the LTE downlink cannot be directly applied to the LTE uplink. Lim et al. [5] showed the NP-hardness of proportional fair packet scheduling for an SC-FDMA system and proposed four heuristic algorithms for the problem. Chang et al. [6] and Chao et al. [7] addressed the robust rate constraint which was not taken into account in many previous studies. In [6], two algorithms were proposed to maximize the sum throughput. In [7], a window-based algorithm was proposed to maximize the total system throughput. In [8], Kwon et al. proposed a QoS uplink scheduling algorithm for LTE in combination with a delay estimation. Kaddour et al. [9] proposed an effective SINR based algorithm for the LTE uplink. Most of the existing works for LTE uplink scheduling have tried to maximize the total throughput. However, without considering the real traffic, many resources are wasted. The results of the above approaches therefore tended to allocate resources to the devices which are not important at all but have good channel conditions. Our method, conversely, will design for surveillance system and the constraints of coverage and quality will both be guaranteed. By considering the security requirements of video surveillance systems (e.g., coverage, monitoring quality), our approach can effectively allocate resources to the crucial devices in the system.

### C. Surveillance system

Research on multi-camera surveillance systems has attracted much attention in recent years. Shen et al. [10] defined Quality of View (QoV), which describes how good the view captured by a camera is by angles and distance between the object and the camera. The joint effect of multiple correlated cameras was studied in [11]. That work aimed to minimize the total number of cameras. In [12], a grid-based flow process of an optimum camera placement algorithm was proposed and the objective was to minimize the overall number of cameras. Dieber et al. [13] focused on camera selection and task assignment in visual sensing network (VSN): under resource limitations, how to set the camera configurations to meet the coverage and QoS requirements was studied. Tseng et al. [14] proposed a k-angle-cover problem for video surveillance with the goal to use the least number of cameras to fulfil the k-angle-cover constraint. In [15], a hybrid scenario of motion sensor and camera was proposed to maximize the system's lifetime, and three objectives were discussed: minimizing energy consumption, maximizing network lifetime, and minimizing monetary cost. Shiang et al. [16] studied how multiple cameras could efficiently share the available wireless network resources and transmit their captured information to a central monitor. It was shown that resource allocation for a wireless surveillance system should take both source characteristics and network conditions into consideration. However, the above approaches were not designed for the SC-FDMA LTE UL. Differing from WiFi surveillance systems and wireless sensor networks (WSN), the resource allocation for LTE is more complex due to its various channel conditions. Although there has been no previous work studied about video surveillance resource allocation in LTE, surveillance systems in LTE have nevertheless recently captured the public's attention.

## III. Background and Formulation

### A. LTE Surveillance Camera

Several surveillance cameras that support LTE have been produced and marketed (e.g. [24], [25]). Features of [24], listed below, will be adopted as our system setting for our simulation. H.264 is used to provide a high compression ratio, and provides three different video resolutions and frame rates, e.g., 720P with 60 fps, and CIF with 30 fps. Thus, according to the video resolution, it has three different bit rate ranges.

### B. LTE UL frame structure

For uplink transmission, LTE employs the Single-carrier Frequency-Division Multiple Access (SC-FDMA) technique. Compared to the Orthogonal Frequency Division Multiple Access (OFDMA) technique used in the LTE downlink, SC-FDMA can transmit a signal with a lower peak-to-average power ratio (PAPR). Thus, using SC-FDMA for LTE UL transmission results in lower power consumption. In SC-FDMA, a basic scheduling unit is called a resource block (RB). In LTE, a radio frame is defined as 10 ms and it can be divided into 10 equal size sub-frames with 1 ms. Each sub-frame is composed of two equal size time slots. Depending on cyclic prefix length, each slot has 6 or 7 symbols. A subcarrier has a 15 kHz bandwidth. One symbol in the time domain, and 1 subcarrier in the frequency domain constitute a resource element (RE). Based on the module and coding scheme (MCS), an RE can carry 2 to 6 bits. An RB is composed of 1 time slot and 12 subcarriers. Thus, an RB has 84 REs with a normal cyclic prefix (7 symbols). A basic time-frequency resource structure of LTE/LTE-A (normal cyclic prefix case) is shown in fig. 1.

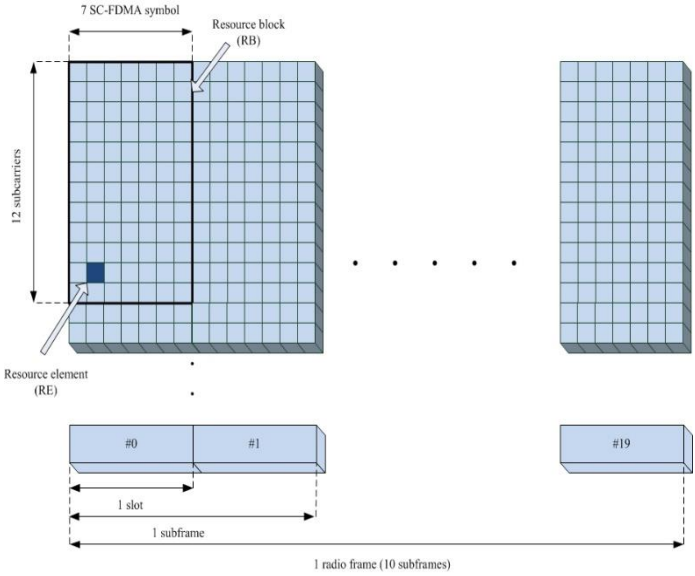

Figure 2. LTE uplink resource structure.

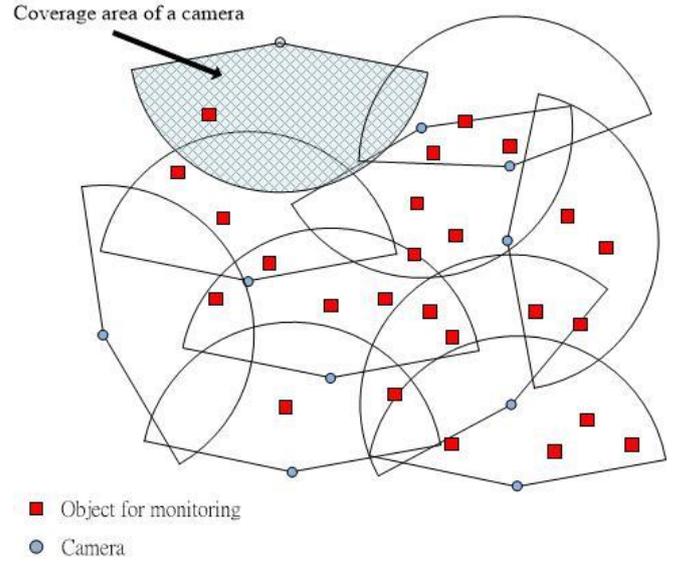

Figure 1. Example of a surveillance system.

C. Problem formulation

For a certain area, an LTE video surveillance system is set for monitoring and multiple cameras are deployed in the system. We assumed that there are total $K$ cameras and $N$ objects for monitoring in the system. The set of $K$ cameras, $\mathcal{K} = \{1, 2, \ldots, K\}$, would upload the video to the remote servers through a base station (e.g., picocell or femtocell) in real time. We assume that the location, sensing area, and transmission range of each camera $k$ are fixed. The set of N objects of interest in the area are denoted as $\mathcal{N} = \{1, 2, \ldots, N\}$. Furthermore, we define a binary indicator $C_{k,n}$ to denote whether the object n is covered by the camera k. $C_{k,n}$ is 1 if the object n is covered by the camera $k$, 0 otherwise. Each camera $k$ has the ability to recognize objects that it covers and the covered objects set of the camera $A_K$ is a subset of $\mathcal{N}$. We define $TP_k$ as the data rate requirement of camera $k$ at each transmission time interval (TTI). When monitoring an object, we usually desire to observe it from a clear view. We define $Q_k$ to represent the monitoring quality value of the camera $k$. In this paper, we adopt the concept of QoV defined in [10] to decide the value of $Q_k$. Thus, we set $Q_k$ as $\sum_{n=1}^{N} \left\{ \omega_\theta \left(1 - \left|\frac{\theta_{kn}}{\pi}\right|\right) + \omega_\phi \left(1 - \left|\frac{2\phi_{kn}}{\pi}\right|\right) + \omega_l \left(1 - \left|\frac{L_{kn}}{L_{Bkn}}\right|\right) \right\}$. In the formula of QoV, $\theta$ is the angle between the subject's body orientation, $\phi$ is the angle between the line passing through the camera center and the horizontal axis, $L_{kn}$ is the distance between the camera $k$ and the object $n$, and $L_{Bkn}$ is the best distance of capturing object $n$ from camera $k$ while $\theta \in (-\pi, \pi]$, and $\phi \in [-\frac{\pi}{2}, \frac{\pi}{2}]$. $\omega_\theta$, $\omega_\phi$, and $\omega_l$ represent the weights of three different metrics. Fig. 2 illustrates an example surveillance scenario in which 10 cameras and 25 monitoring targets are deployed in the area.

In an LTE video surveillance system, the amount of resources for transmitting is limited and may be insufficient. Suppose that there are a total W RBs and they are equally divided into M sub-bands for the LTE uplink. Therefore, each sub-band has $W_m$ RBs and each RB can only be allocated to one camera $k$ at one TTI. We assume that the channel fading for a camera on a sub-band is flat. Based on sounding reference signal (SRS) reported by each camera, each camera $k$ can apply only one proper MCS level in one sub-band $m$ at one TTI and we denote it as $s_{m,k}$. Moreover, we define $r_{m,k}$ to represent the number of the required RBs of camera $k$ when it uses RBs in sub-band m for data transmission which is calculated by $TP_k$ and $s_{m,k}$. We further let $x_{m,k} \in \{0,1\}$ be the allocation indicator; $x_{m,k}$ will be 1 if the RBs are allocated to the camera k in sub-band m, 0 otherwise.

We define our problem as follows: if there are M sub-bands and only $W_m$ RBs available in each sub-band for the LTE M2M surveillance system, how can we select $\mathcal{K}'$ cameras from $\mathcal{K}$ and how can we allocate $W_m$ RBs in each sub-band to the selected cameras in order to make all the N objects that are covered by the selected cameras reach the throughput requirement for each selected camera, and maximize the monitoring total quality of the system.

The problem is formulated as the following objective function

$$\text{maxmize } z = \sum_{m=1}^{M} \sum_{k=1}^{K} x_{m,k} \cdot Q_k$$

which is subject to the following constraints:

$$\sum_{m=1}^{M} x_{m,k} \leq 1, \quad \forall k \in \mathcal{K} \quad (1)$$

$$\sum_{m=1}^{M} \sum_{k=1}^{K} C_{k,n} \cdot x_{m,k} > 0, \quad \forall n \in \mathcal{N} \quad (2)$$

$$\sum_{k=1}^{K} r_{m,k} \cdot x_{m,k} \leq W_m, \quad \forall m \in \{1, 2, \ldots, M\} \quad (3)$$

$$x_{m,k} \in \{0, 1\}, \forall k \in \mathcal{K}, \forall m \in \{1, 2, \ldots, M\} \quad (4)$$

Constraint (1) shows that each camera can only transmit data in one sub-band at any one time. Constraint (2) means that all objects should be covered by cameras which will be allocated resources for transmission. Monitoring all objects is necessary for the system. Constraint (3) indicates that the amount of RB requirement in one sub-band cannot exceed the total number of RBs in a sub-band. Furthermore, when each RB is allocated to a camera, it should be allocated in a continuous manner to follow the contiguous constraints of SC-FDMA.

## IV. PROPOSED ALGORITHM

First, we describe a baseline algorithm based on SNR measurements. Then, a small scale example is given to explain why this approach is not suitable for a surveillance system in the LTE UL. Next, we propose our algorithm which includes two parts: 1) Monitoring Quality Based Scheduling (MQBS), and 2) dynamic allocation. MQBS aims to maximize the performance of the system while dynamic allocation adjusts the allocation of cameras when background traffic arrives. After MQBS is performed, the system has a basic allocation map of the cameras. When a camera is served in the system, it should be served for a period of time to provide a stable video. Thus, we do not want the system serving different cameras at each TTI. MQBS is not performed at each TTI but performed periodically (e.g. 10 seconds). However, the network condition may change with time. Some new background traffic may arrive while others may depart. For arrival traffic, dynamic allocation changes the allocated cameras or the allocated resources to meet both the coverage constraint and performance. For departure traffic, although available resources are increased when background traffic departs, in order to prevent modifying the system too frequently, the system would not serve a new camera.

### A. Baseline algorithm

The main propose of most scheduling algorithms for LTE resource allocation is to maximize the total network throughput. These kinds of algorithms adopt the measured SNR value or the channel quality feedback as comparison criteria. Based on the measured SNR value or the channel quality feedback, the scheduler selects a proper MCS for the UE. MCS is used to determine the likely data rate of the UE in the RB. Since each UE has its own data rate requirement, the UE with a higher MCS level needs fewer RBs to transmit data. When more UEs are served in the system, a higher network throughput is reached.

Therefore, assigning the RB to the camera with the best MCS level is the most intuitive approach. In each iteration, the algorithm allocates resources to an unscheduled camera which has the best MCS level and covers objects that are not yet covered. After all objects are covered by the selected cameras, the algorithm will iteratively allocate resources to an unscheduled camera with the best MCS level until no more cameras can be served in the system.

We provide an example to explain why the baseline scheme cannot work well in the proposed system. The scenario in this example is a multi-camera surveillance system with seven cameras and six observation targets. The available network resources are divided into three sub-bands and each sub-band has five RBs. A camera can only use RBs in one sub-band and has the same MCS level in the same sub-band. Each camera has its own coverage set and monitoring quality. To simplify the example, we suppose that the data rate of cameras is the same and we use the required RBs directly instead of using the MCS level. The information of each camera is depicted in Table. 1(a).

We use the element $r_{j,k}$, as the RB requirement of the camera $k$ on sub-band $j$. Using the baseline scheme as the sample scheduling algorithm, the first smallest RBs requirement, $r_{1,6}$, is 2. The scheduler allocates RB1 and RB2 in sub-band1 to camera6. Next, given that $r_{2,1}$ is the smallest, the scheduler allocates RB1, RB2 and RB3 in sub-band2 to camera1. The scheduler then follows the same procedure and allocates resources to camera2 and camera4. After all objects and no more cameras can be served in the system, the total quality that provided by the served camera is 15. The result of the baseline algorithm is shown in Table. 1(b). In this example, the baseline algorithm does not work well because it only takes channel quality into account. Some important cameras may not be scheduled due to its poor channel quality.

### B. Monitoring Quality Based Scheduling (MQBS)

A greedy algorithm for uplink resources allocation will be proposed in this part. Our objective is to maximize the total monitoring quality under spectrum and coverage constraints. Thus, two criteria, coverage and monitoring quality, are important to the algorithm that affects the system. In our approach, the first step selects the cameras to fulfill the coverage

Table 1. Scheduling example for a multi-camera surveillance system.

|  | Sub-band1 | Sub-band2 | Sub-band3 | Coverage | Quality |
|---|---|---|---|---|---|
| Camera1 | 5 | 3 | 5 | {2,5} | 4 |
| Camera2 | 4 | 5 | 3 | {1,2,4} | 5 |
| Camera3 | 4 | 4 | 4 | {1,4,5} | 7 |
| Camera4 | 3 | 4 | 3 | {5,6} | 3 |
| Camera5 | 4 | 3 | 5 | {2,3,4} | 6 |
| Camera6 | 2 | 2 | 2 | {1,3} | 3 |
| Camera7 | 4 | 4 | 4 | {4,5,6} | 5 |

(a) Channel condition, coverage set, and quality of each camera.

|  | RB1 | RB2 | RB3 | RB4 | RB5 |
|---|---|---|---|---|---|
| Sub-band 1 | Camera6 | Camera6 | Camera4 | Camera4 | Camera4 |
| Sub-band 2 | Camera1 | Camera1 | Camera1 |  |  |
| Sub-band 3 | Camera2 | Camer2 | Camera2 |  |  |

(b) Scheduling result of the baseline.

|  | RB1 | RB2 | RB3 | RB4 | RB5 |
|---|---|---|---|---|---|
| Sub-band 1 | Camera3 | Camera3 | Camera3 | Camera3 |  |
| Sub-band 2 | Camera5 | Camera5 | Camera5 | Camera6 | Camera6 |
| Sub-band 3 | Camera7 | Camera7 | Camera7 | Camera7 |  |

(c) Scheduling result of the proposed method.

constraint, while the second step utilizes the remaining resources to approximate our objective.

*1) Coverage assurance phase*

In this phase, we make sure that every object is being monitored by at least one camera. Before the main procedure of the algorithm, the scheduler first initializes monitoring quality $Q_k$, coverage indicators $C_{k,n}$ and RB requirement values $r_{m,k}$ for each camera. The RB requirement values $r_{m,k}$ are calculated by using SRS report and the data rate requirement $TP_k$. In order to guarantee the coverage, the scheduler first iteratively examines each object. If object $n$ has not yet been covered by the selected cameras, the scheduler chooses the camera $k$ that covers the object $n$ with largest monitoring quality $Q_k$ to be served in the system. Since $Q_k$ is defined as an aggregation of monitoring quality for each object provided by camera $k$, a camera $k$ with higher $Q_k$ tends to have a larger coverage set. Thus, using $Q_k$ directly as a comparison metric can reduce the amount of uncovered objects and simultaneously approximate our objective. After the camera $k$ is selected, by using $r_{m,k}$, a sub-band $m$, in which the selected camera $k$ needs the fewest RBs when transmitting data on the sub-band, will be chosen. More remaining resources are available to serve other cameras by using the fewest RBs. After allocating the resources to the camera $k$, remaining number of RBs in $m$ is re-calculated. The allocation map $x_{m,k}$ is then set to be 1. Finally, due to a new camera being served, the scheduler updates the coverage area to understand which objects are now covered. The procedure is repeated until all the objects are covered.

*2) Monitoring quality improvement phase*

In this phase, we improve the performance of the system. Since there may be remaining RBs in the system, more cameras can be served to improve the monitoring quality. First, the scheduler sorts out the cameras that have not been selected by $Q_k$ in descending order into a list. Then, the scheduler iteratively checks whether or not it is possible to serve the camera. If there are sub-bands $m$ that can serve the camera $k$, a sub-band with the lowest $r_{m,k}$ will be selected. Allocation map $x_{m,k}$ and remaining result are updated after a new camera is served.

Our approach can obtain a higher monitoring quality than that of the baseline algorithm. In this example, we first perform the coverage assurance phase. Camera2, camera3, and camera6 all cover object1, but camera3 has the best Quality. Considering the RB requirement of each sub-band, sub-band1 is selected to serve camera3. After re-calculating the coverage, the scheduler should next select a camera for object2. As in the procedure above, camera5 and camer7 are selected and are served in the most appropriate sub-band. After the coverage constraint is met, the system will perform our quality improvement phase. The scheduler examines the cameras after sorting the remaining cameras. Although camera6 has the lowest quality, it can be served in the system. The scheduler finally selects camera3, camera5, camera6, and camera7 to be served. The total quality is 21, which outperforms the result of the baseline algorithm.

### C. The dynamic allocation

To adapt to the network condition quickly, we should avoid re-calculating the whole allocation map. This means that we should

---

**Algorithm 1: MQBS Coverage assurance phase**

1: **Input** $Q_k, C_{k,n}, r_{m,k}, R_m$
2: $G \leftarrow \emptyset, X_{m,k} \leftarrow 0$  //G is covered objects set
3: **for** $n = 1$ to $N$ **do**
4:   **if** $n \notin G$ **then**
5:     $v \leftarrow 0$
6:     **for** $k = 1$ to $K$ **do**
7:       **if** $C_{k,n} = 1$ and $v < Q_k$ **then**
8:         $CAM \leftarrow k$
9:         $v \leftarrow Q_k$
10:   $r_{min} = \infty$
11:   **for** $m = 1$ to $M$ **do**
12:     **if** $r_{m,CAM} < r_{min}$ and $r_{m,CAM} \leq R_m$ **then**
13:       $r_{min} \leftarrow r_{m,CAM}$
14:       $SB \leftarrow m$
15:   $X_{SB,CAM} \leftarrow 1$
16:   $R_{SB} \leftarrow R_{SB} - r_{SB,CAM}$
17:   **for** $n = 1$ to $N$ **do**
18:     **if** $C_{CAM,n} = 1$ **then**
19:       $G \leftarrow G \cup \{n\}$
20: **return** $X$

---

**Algorithm 2: MQBS Monitoring quality improvement phase**

1: **Input** $Q_k, r_{m,k}, R_m, X_{m,k}$
2: sort remaining cameras into list T by $Q_k$ in descending order
3: **while** $T \neq \emptyset$
4:   $CAM \leftarrow$ first element of $T$
5:   $SEL \leftarrow$ FALSE
6:   $r_{min} \leftarrow \infty$
7:   **for** $m = 1$ to $M$ **do**
8:     **if** $r_{m,CAM} < r_{min}$ and $r_{m,CAM} \leq R_m$ **then**
9:       $r_{min} \leftarrow r_{m,CAM}$
10:       $SB \leftarrow m$
11:       $SEL \leftarrow$ TRUE
12:   **if** $SEL =$ TRUE **then**
13:     $X_{SB,CAM} \leftarrow 1$
14:     $R_{SB} \leftarrow R_{SB} - r_{SB,CAM}$
15:   $T \leftarrow T - \{CAM\}$
16: **return** $X$

---

use a simpler way to adjust the network. The MQBS is performed periodically but not at every TTI. In this part, we will propose an algorithm for adapting the network condition when background traffic arrives. The proposed algorithm first decide the transmission sub-band for the new arrival background traffic. Then, to adapt the traffic load, the camera allocation map will be modified slightly. In the algorithm, two actions, 1) re-routing a camera, 2) removing a camera, will be applied. Both actions offload the load of the sub-band that has a high load under the constraint of coverage.

*1) RA for background traffic and offloading decision phase*

The main work of this phase is to allocate resources for new arrival background traffic and decide whether to offload the current network. At first the scheduler calculates the current remaining RB of each sub-band. The scheduler then selects a candidate sub-band for the new arrival background traffic with the smallest (RB requirement/remainin g RB) in the sub-band. The purpose of using the comparison metric, (RB requirement/ remaining RB), is to serve the traffic in a low load sub-band. Unfortunately, if the RB requirement is larger than remaining RB of the candidate sub-band, there is no available sub-band for the new arrival traffic. The main idea is to offload the candidate sub-band to a low load sub-band. The scheduler recalculates the remaining RB of the candidate sub-band. If the remaining RB is smaller than the $th_h$, an offloading decision is needed. The scheduler then selects a sub-band with the largest remaining RB for offloading. If the remaining RB of the selected sub-band is larger than the $th_L$, the algorithm enters *re-routing phase*, otherwise, enters *removing phase*. Both $th_h$ and $th_L$ are thus offloading decision thresholds.

*2) Re-routing phase*

In this function, the scheduler re-routes a camera from the high load candidate sub-band into a low load sub-band. First, the scheduler chooses the camera in the candidate sub-band with the highest RB requirement to be the re-routing target. Since the coverage constraint has to be followed, the scheduler calculates the coverage set when the re-route target is not in the system. From the unscheduled cameras, a camera which can cover all uncovered objects and has the smallest RB requirement in the offloading sub-band will be selected as a new scheduled camera. The re-routing procedure is only performed when the remaining RB of the offloading sub-band is still under the $th_h$. With the limitation of $th_h$, the system will not suffer a ping-pong effect between sub-bands.

*3) Removing phase*

This function is only performed when all sub-bands are crowded. A camera in the high load candidate sub-band will be removed from scheduling while all objects are still covered by scheduled cameras. First, the scheduler calculates the coverage set from cameras that are served in the sub-bands except cameras in the candidate sub-band. After this procedure, the scheduler can know the coverage set without the cameras in the candidate sub-band. The scheduler then sorts the cameras in the candidate sub-band by the amount of the coverage objects in descending order. A camera which covers large objects amount usually has a great influence on the coverage set. According to the sorted list, the scheduler iteratively checks if the camera is needed to meet the coverage requirement. If the camera is required, it is marked and renewed in the coverage set. Finally, a camera that is not required and has the least monitoring quality is removed from the scheduling.

D. *Complexity analysis*

In the coverage assurance phase of MQBS, iterations of the main assignment procedure will be performed after initializations. Each iteration includes 1) a camera with highest monitoring quality, 2) a sub-band in which the camera has the lowest RB requirement, and 3) an updating coverage area with the complexity of $\mathcal{O}(K)$, $\mathcal{O}(M)$, and $\mathcal{O}(N)$ respectively. Thus, the complexity for the coverage assurance phase is $\mathcal{O}(N(K + M + N))$. The monitoring quality improvement phase will be performed right after coverage assurance phase. A sorting for the cameras by quality will be performed first, and it has complexity $\mathcal{O}(K \log K)$. According to the sorting result, the algorithm iteratively checks the cameras to see whether or not each camera can be served and assign a best sub-band to it if it can be served. The complexity of this procedure is $\mathcal{O}(KM)$. The total complexity of the quality improvement phase is $\mathcal{O}(K(\log K + M))$. The total complexity for MQBS is $\mathcal{O}(NK + NM + N^2 + K \log K + KM)$.

---

**Algorithm 3: RA for background traffic and offloading decision phase**

1: **Input** $R_m$, $r_{m,arrival}$
2: $w \leftarrow \infty$
3: **for** $m = 1$ to $M$ **do**
4:   **if** $R_m > r_{m,arrival}$ **then**
5:     $w_m \leftarrow r_{m,arrival}/R_m$
6:     **if** $w_m < w$
7:       $w \leftarrow w_m$
8:       $CandidateM \leftarrow m$
9: **if** $R_{CandidateM} \leq threshold_h$
10:   $load \leftarrow 0$
11:   **for** $m = 1$ to $M$ **do**
12:     **if** $R_m \geq threshold_L$ && $R_m > load$
13:       $load \leftarrow R_m$
14:       $offloadM \leftarrow m$
15:   **if** offloadM is not null **then**
16:     offload by re-routing a camera
17:   **else**
18:     offload by removing a camera

---

**Algorithm 4: re-routing phase**

1: **Input** $R_m$, $r_{m,k}$, $C_{k,n}$, CandidateM, offloadM
2: $r_{min} \leftarrow 0$, $G \leftarrow \emptyset$
3: **for** $k$ in *CandidateM* **do**
4:   **if** $r_{CandidateM,k} > r_{min}$ **then**
5:     $r_{min} \leftarrow r_{CandidateM,k}$
6:     $kRemove \leftarrow k$
7:     $kJoin \leftarrow k$
8: **for** camera $k$ that is scheduled **do**
9:   **for** $n = 1$ to $N$ **do**
10:     **if** $C_{k,n} = 1$ **then**
11:       $G \leftarrow G \cup \{n\}$
12: $r_{max} \leftarrow \infty$
13: **for** camera $k$ that is unscheduled **do**
14:   $CAM \leftarrow TRUE$
15:   **for** $n = 1$ to $N$ **do**
16:     **if** $n \notin G$ and $C_{k,n} = 0$ **then**
17:       $CAM \leftarrow FALSE$
18:       Break
19:   **if** $CAM = TRUE$ and $r_{offloadM,k} < r_{max}$
20:     $kJoin \leftarrow k$
21:     $r_{max} = r_{offloadM,k}$
22: **if** $R_{offloadM} - r_{max} > threshold_h$
23:   return $kRemove$, $kJoin$
24: **else**
25:   return null

In the first phase of dynamic allocation, we need to 1) decide the sub-band for the new arrival background traffic, and 2) decide the sub-band for offloading, and the complexity of both parts is $O(M)$. In the re-routing phase, selecting a camera to be re-routed is $O(K)$, and re-calculating the coverage is $O(KN)$. The total complexity of re-routing phase is $O(KN)$. In the removing phase, $O(KN)$ is needed in order to know the coverage provided by other sub-bands. Then a sorting with complexity $O(KlogK)$ is performed. To know which cameras is still necessary is $O(KN^2)$. Finally, a camera which is not necessary and has the worst quality is removed. The complexity of this procedure is $O(K)$. Therefore, the complexity for the removing phase is $O(K(logK + N^2))$. The total complexity of dynamic allocation is $O(M + K(logK + N^2))$.

## V. SIMULATION

In this section, we will simulate our method and compare the result to the results of other scheduling methods in the LTE uplink.

### A. Simulation setting

To simulate a video surveillance system, we consider a monitored area is a 2-D circle plane with a radius of 250m and a base station at the center of the area. Thus, the area enclosed by the circle is $250 \times 250 \times \pi \ m^2$. The area can be considered as a school, a shopping mall, a subway station, or any other important space in a city. In the area, $K$ cameras and $N$ objects are uniformly-distributed and scattered over the whole area of the cell. After we generate the input pair of the objects set and the cameras set, we check whether all objects in the objects set can be covered by cameras in the cameras set to make sure the coverage constraint can be fulfilled. Otherwise, using the input pair which the coverage constraint cannot be met will lead to an error in the results.

The system bandwidth is set at 10 MHz. By [18], an RB has a 180kHz bandwidth. In a 10MHz bandwidth, there are a total of 50 RBs in a time slot. However, some of RBs will be used for control signals in reality. We suppose that there are a total of 48 available RBs in the system for data transmission and those RBs are equally separated into sub-bands. For a realistic simulation, the modulation coding scheme is based on [19], and the path loss and shadowing model are based on [20]. We suppose that the base station is surrounded by 6 other base stations. Thus, the system will suffer from the inter-cell interference. The transmission power of cameras is set to be 24dbm.

In the simulation, a spectrum efficient greedy-based method (G-B) [17] is chosen for comparison. Under different simulation settings and inputs, we compare MQBS to this spectrum efficient greedy-based method. Because the G-B method is proposed without considering a video surveillance application, we have to examine the covered objects by the served cameras to make sure that the results satisfy the coverage constraint. In addition, we also found the optimal solution with the proposed Integer Linear Programming formulation which was solved by Gurobi [23]. In each experiment, each case is run at least 500 times. The system settings are summarized in table 2.

The following parameters will be used to evaluate our proposed method: 1) number of objects, 2) number of cameras, 3) distance of view, and 4) angle of view. To evaluate the performance, the monitoring quality of camera k (i.e. $Q_k$) is set as $\sum_{n=1}^{N}\left\{\left(1-\left|\frac{\theta_{kn}}{\pi}\right|\right) + \left(1-\left|\frac{L_{kn}}{L_{Bkn}}\right|\right)\right\}$. The two metrics of RB requirement and quality of the total system are used for performance comparison.

```
Algorithm 5: removing phase
1:  Input C_{k,n}, CandidateM
2:  G ← ∅
3:  for camera k that is scheduled without in CandidateM
    do
4:    for n = 1 to N do
5:      if C_{k,n} = 1 then
6:        G ← G ∪ {n}
7:  Sort cameras in CandidateM into list T by the amount
    of the coverage objects in descending order
8:  for n = 1 to N do
9:    if n ∉ G then
10:     for k in T do
11:       if C_{k,n} = 1
12:         G ← G ∪ {n}
13:         Req_k = TRUE
14:         for n = 1 to N do
15:           if C_{k,n} = 1 then
16:             G ← G ∪ {n}
17:         Break
18: v ← ∞
19: for k in T do
20:   if Req_k = FALSE && Q_k < v then
21:     kRemove ← k
22:     v ← Q_k
23: return kRemove
```

Table 2. Simulation Setting

| Parameter | Setting |
| --- | --- |
| System bandwidth | 10MHz |
| Uplink bandwidth per RB | 180kHz |
| Total available RBs # | 48 |
| Sub-carriers per RB | 12 |
| Cyclic prefix | 7 |
| Sub-band # | 4 |
| Scheduling time interval | 1ms (1 sub-frame) |
| Modulation coding scheme | QPSK 1/3, 1/2, 2/3, 3/4 16QAM 1/2, 2/3,3/4 |
| Monitoring area | A circle with radius 250m |
| Transmission power | 24dBm |

### B. Simulation results

We first investigate the impact of object amount to the system by varying the object amounts from 30 to 80 in the system. We set the total camera amount to 50, angle of view to 150 degrees, and distance of view to 100m as the default setting. Objects for monitoring are randomly distributed in the circle area with a 250m radius. In fig. 3, we can see that our method outperforms G-B. Fig. 3(a) shows the minimum RB requirement for covering all objects, and it is clear that both methods need more RBs when the object amount increases. However, our method always requires fewer RBs to cover all objects. Generally, the RB requirement highly depends on the object amount because the

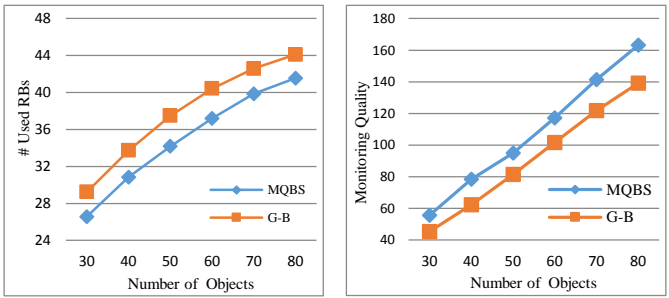

(a) Minimum number of RBs for constraint requirements

(b) Monitoring quality when using minimum number of RBs

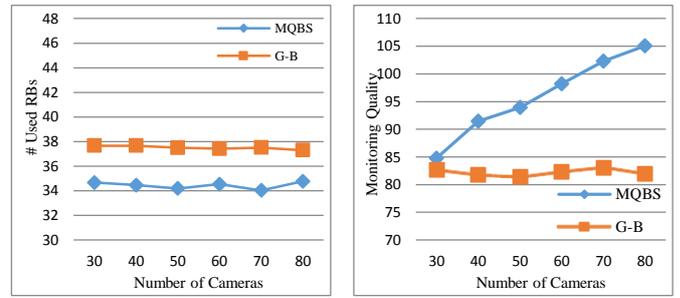

(a) Minimum number of RBs for constraint requirements

(b) Monitoring quality when using minimum number of RBs

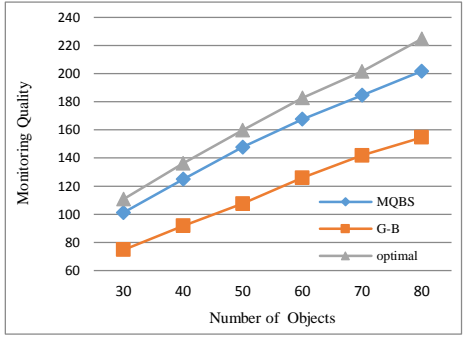

(c) Maximum monitoring quality if all RBs are available

Figure 3. Performance comparison of two approaches under various number of objects.

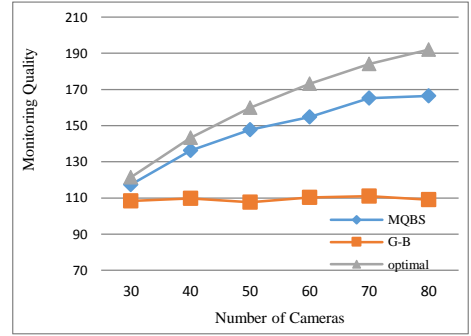

(c) Maximum monitoring quality if all RBs are available

Figure 4. Performance comparison of two approaches under various number of cameras.

system needs to serve more cameras to fulfill the coverage constraints. Fig. 3(b) shows the sum of quality of the cameras that are selected for fulfilling the coverage constraint. From figs. 3(a) and 3(b), we can see that our method needs fewer RBs while simultaneously being able to achieve a higher quality. The total quality of the system when all RBs are available for the system is depicted in fig. 3(c). It is shown that our method is better than the G-B and, moreover, the gap between our method and G-B increases when the amount of the objects grows.

We next explore the influence of the camera amount on the system in fig. 4. We vary the camera amount from 30 to 80 while keeping fixed the object amount at 50, angle of view at 150 degrees, and distance of view at 100m. Our method requires fewer RBs than G-B, as shown in fig. 4(a). The RB requirement values of both methods remain nearly the same value when camera amount grows. This shows that no matter how many cameras are in the system, the system only needs to serve nearly the same amount of cameras to fulfill the coverage constraint. However, the quality of monitoring is different. From fig. 4(b) and 4(c), it is evident that the difference of quality between the two methods gets larger when the camera amount grows. This is because there are more choices when camera diversity becomes higher. Our method make better choices and, thus, the quality gap becomes larger.

In fig. 5, we present the effect of angle of view. We set the object amount to 50, camera amount to 50 and distance of view to 100m. The angle of view is varied from 90 to 180 degrees. The RB requirements of both methods decrease when angle of view becomes larger, yet our method always requires fewer RBs than G-B as depicted in fig. 5(a). Since one camera has higher coverage ability when the angle of view increases, fewer cameras needed to be served to ensure coverage. In other words, the results show that more cameras are unnecessary for uploading the captured video. From fig. 5(b) and 5(c), the total monitoring quality provided by our method are also better than those for G-B. As the angle of view becomes larger, the average monitoring quality of a camera also becomes larger. Moreover, more cameras can be served to improve the quality since there is a lower minimum RB requirement for coverage. Both of these contribute to the result of increased quality, as shown in fig. 5(c).

In fig. 6, we investigate the effect of the view distance on the system by setting the distance of view from 80 to 140m. We set both camera amount and object amount to 50 and angle of view to 150 degrees. As can be seen in fig. 6, both minimum RB requirement and total quality of the system of our method outperform those of G-B. The reason for the decreasing RB requirement is the same as for the angle of view, namely there is a higher coverage ability of cameras. Moreover, a higher coverage ability also results in an increase in total quality.

## VI. CONCLUSIONS

The scheduling algorithm of resource allocation plays an important role for LTE. An efficient scheduling will greatly improve the system capacity of LTE. Thus, most of previous works about LTE UL resource allocation took channel quality into account to maximize the system capacity. However, there is no scheduler which takes into consideration the demand for a specific service. Due to the proliferation of M2M applications, a scheduler which is designed for a specific application is crucial in the future. In this paper, we investigated the UL resource allocation problem for the LTE video surveillance system. Addition to the constraints of LTE UL, we further took both the coverage requirements and camera monitoring qualities into consideration, and formulated the scheduling problem to

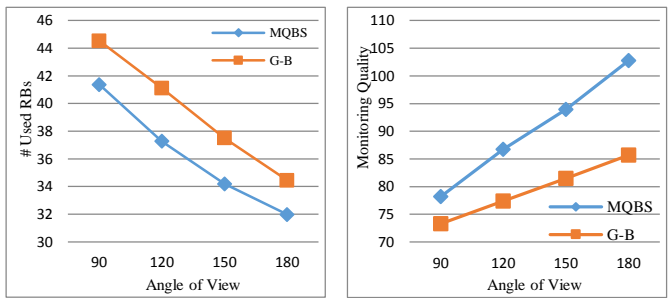
(a) Minimum number of RBs for constraint requirements
(b) Monitoring quality when using minimum number of RBs

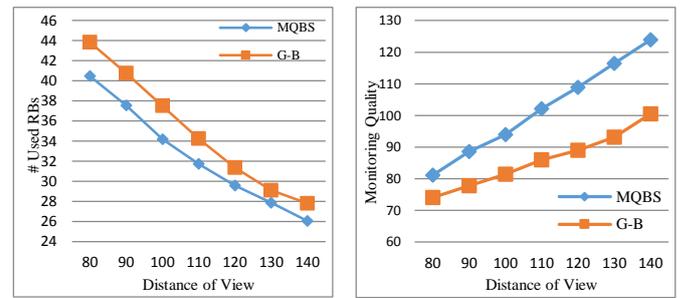
(a) Minimum number of RBs for constraint requirements
(b) Monitoring quality when using minimum number of RBs

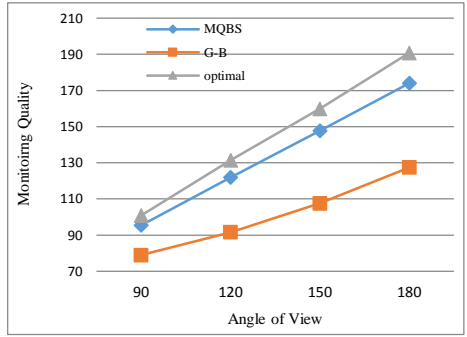
(c) Maximum monitoring quality if all RBs are available

Figure 5. Performance comparison of two approaches under various values of angle of view.

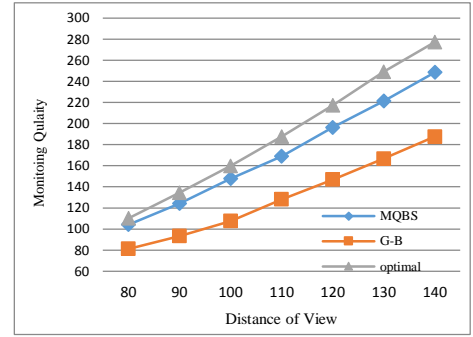
(c) Maximum monitoring quality if all RBs are available

Figure 6. Performance comparison of two approaches under various values of distance of view.

maximize the total system monitoring quality. We studied a baseline scheduling algorithm based on SNR first and showed that it does not perform well for LTE surveillance system. To solve the problem, we proposed a heuristic algorithm (MQBS) based on monitoring quality. Moreover, a heuristic offloading method when background traffic arrives was also presented. The simulation results demonstrate that the minimum RB requirement can be decreased while the total monitoring quality of the surveillance system can be increased in comparison with an existing approach for LTE networks.